\documentclass[final,3p,times,twocolumn]{elsarticle}

\usepackage{epsfig}

\usepackage{graphicx}%
\usepackage{multirow}%
\usepackage{amsmath,amssymb,amsfonts}%
\usepackage{amsthm}%
\usepackage{mathrsfs}%
\usepackage[title]{appendix}%
\usepackage{xcolor}%
\usepackage{textcomp}%
\usepackage{manyfoot}%
\usepackage{booktabs}%
\usepackage{graphicx} 
\usepackage{algorithmicx}%
\usepackage{algpseudocode}%
\usepackage{listings}%
\usepackage{multirow}%
\usepackage{subfig}
\usepackage{tabularx}
\usepackage{url}
\usepackage{booktabs} 
\usepackage[ruled,vlined]{algorithm2e}
\usepackage{rotating} 
\usepackage{float}

\begin{document}

\begin{frontmatter}



\title{Efficient Breast and Ovarian Cancer Classification via ViT-Based Preprocessing and Transfer Learning}

\author[aff1]{Richa Rawat}
\ead{richa.rawat2@uta.edu}

\author[aff2]{Faisal Ahmed\corref{cor1}}
\ead{ahmedf9@erau.edu}

 \cortext[cor1]{Corresponding author}

\address[aff1]{Department of Mathematics, University of Texas at Arlington, 701 S Nedderman Dr, Arlington, 76019, Texas, USA}

 \address[aff2]{Department of Data Science and Mathematics, Embry-Riddle Aeronautical University, 3700 Willow Creek Rd, Prescott, Arizona 86301, USA}

\begin{abstract}
Cancer is one of the leading health challenges for women, specifically breast and ovarian cancer. Early detection can help improve the survival rate through timely intervention and treatment. Traditional methods of detecting cancer involve manually examining mammograms, CT scans, ultrasounds, and other imaging types. However, this makes the process labor-intensive and requires the expertise of trained pathologists. Hence, making it both time-consuming and resource-intensive. In this paper, we introduce a novel vision transformer (ViT)-based method for detecting and classifying breast and ovarian cancer. We use a pre-trained ViT-Base-Patch16-224 model, which is fine-tuned for both binary and multi-class classification tasks using publicly available histopathological image datasets. Further, we use a preprocessing pipeline that converts raw histophological images into standardized PyTorch tensors, which are compatible with the ViT architecture and also help improve the model performance.

We evaluated the performance of our model on two benchmark datasets: the BreakHis dataset for binary classification and the UBC-OCEAN dataset for five-class classification without any data augmentation. Our model surpasses existing CNN, ViT, and topological data analysis-based approaches in binary classification. For multi-class classification, it is evaluated against recent topological methods and demonstrates superior performance. Our study highlights the effectiveness of Vision Transformer-based transfer learning combined with efficient preprocessing in oncological diagnostics.

\end{abstract}





\end{frontmatter}



\section{Introduction}\label{sec1}
Breast and ovarian cancer are among the most common forms of cancer and are life-threatening for women worldwide \cite{biom15050657}. In 2020, approximately 2.3 million new cases were reported, and this number is expected to keep rising over time \cite{sung2021global, mccormack2023current}. According to research conducted by the World Health Organization (WHO), ovarian cancer ranks fifth in terms of cancer-related deaths. Initially, ovarian cancer was predominantly seen in older women, but changing lifestyles have led to an increase in cases among younger women. Therefore, it is crucial to develop tools for early diagnosis \cite{j2024worldwide}. Early and accurate detection of cancer can significantly improve survival rates by enabling patients to receive timely and appropriate treatment. Current diagnostic methods require meticulous manual examination of various imaging types, including mammograms, CT scans, and ultrasounds. However, these assessments can be labor-intensive, subjective, and may vary greatly between different evaluators. Such challenges can lead to delays in clinical decision-making, ultimately affecting patient outcomes. Thus, there is a pressing need for advancements in diagnostic technologies to streamline and enhance early cancer detection \cite{gurcan2009histopathological}.

In recent years, several deep learning methods have been introduced to improve early detection and classification in medical imaging. Among these, convolutional neural networks (CNNs) have demonstrated significant effectiveness in analyzing various imaging modalities, including mammograms \cite{shenMammography}, sonograms \cite{Byrasonnography, gao2022deep}, histopathological slides \cite{breasthist, wang2022weakly}, and MRIs \cite{saida2022diagnosing}, particularly for the detection of breast and ovarian cancers.
However, CNNs have some drawbacks, including the need for large volumes of labeled training data and substantial computational memory requirements. In addition, CNNs struggle to capture complex and long-range spatial dependencies in medical images effectively \cite{Alexchallenges}.

Transformers are deep learning models that were originally developed for natural language processing. Their self-attention mechanism enables them to process long sentences and understand the context and relationships between nearby and distant words \cite{vaswani2023attentionneed}. Recently, transformers have also been successfully applied to computer vision tasks, leading to the development of vision transformers (ViTs) \cite{DingTransformers, VisionEmerald}. These vision transformers have also shown promise in classification tasks related to breast and ovarian cancer \cite{LaidTokenMixer}.

This paper presents a methodology based on Vision Transformers (ViT) for the automated diagnosis of breast and ovarian cancers using publicly available datasets. The approach uses transfer learning with a pre-trained ViT model \cite{dosovitskiy2020image}, which is fine-tuned for binary and five-class classification tasks using the Breast Cancer Histopathological Image Classification (BreakHis) dataset and the University of British Columbia Ovarian Cancer Subtype Classification and Outlier Detection (UBC OCEAN) dataset \cite{UBC-OCEAN}. A pipeline has been developed to optimize the ViT architecture, enhancing computational efficiency and robustness. Experimental results demonstrate that the proposed model achieves state-of-the-art performance in classifying breast and ovarian cancers.

\begin{figure}[t!]
	\centering
	\subfloat[\scriptsize Benign image sample at 40x magnification.\label{fig:B_40-sample1}]{%
		\includegraphics[width=0.2\linewidth]{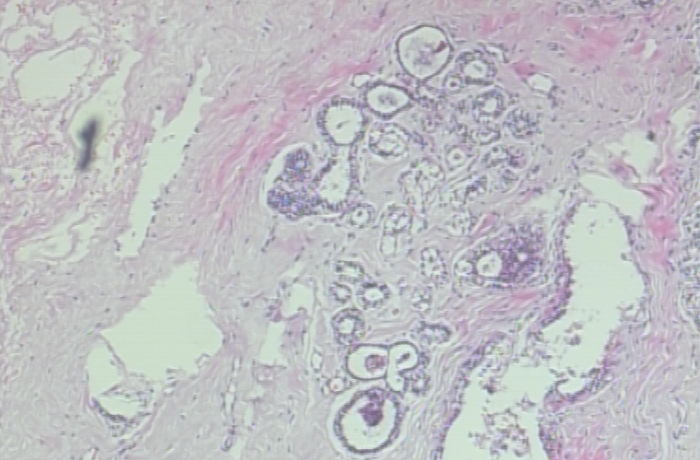}}
	\hfill
	\subfloat[\scriptsize Benign image sample at 100x magnification.\label{fig:B_100-sample2}]{%
		\includegraphics[width=0.2\linewidth]{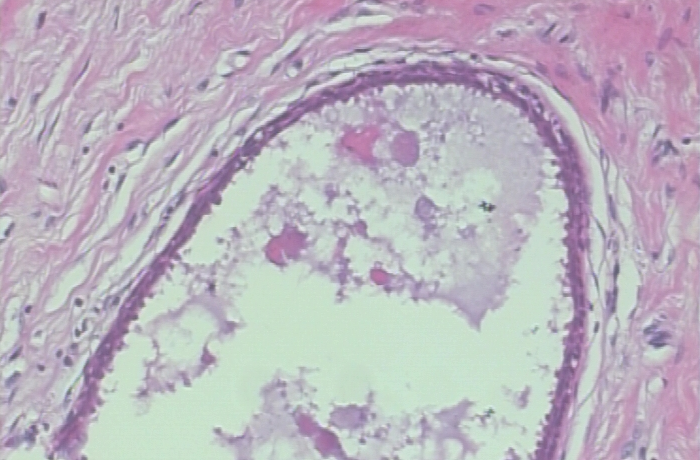}}
	\hfill
	\subfloat[\scriptsize Benign image sample at 200x magnification.\label{fig:B_200-sample3}]{%
		\includegraphics[width=0.2\linewidth]{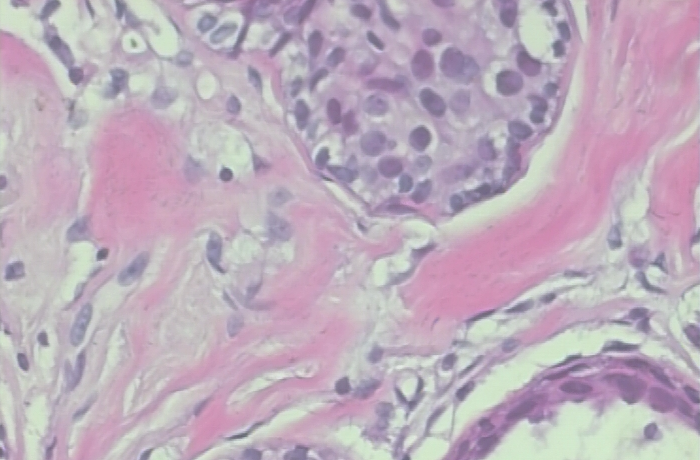}}
    \hfill
	\subfloat[\scriptsize Benign image sample at 400x magnification.\label{fig:B_400-sample2}]{%
		\includegraphics[width=0.2\linewidth]{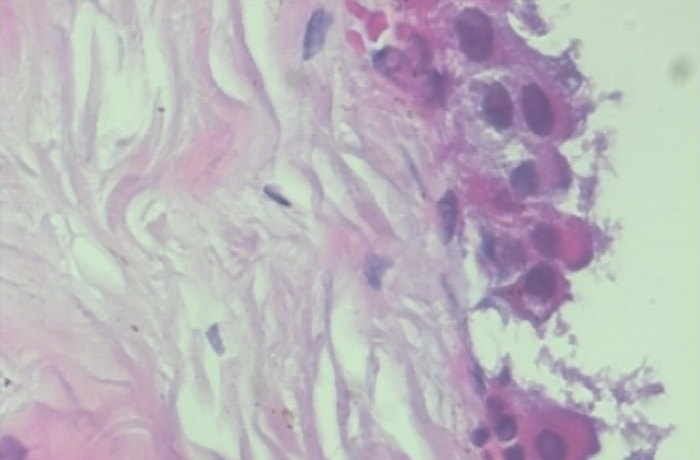}}

	\caption{\footnotesize Representative benign tissue image samples of the BreakHis dataset at different magnifications (40x, 100x, 200x, and 400x) for visual comparison.}
	\label{fig:b-image-samples}
\end{figure}

\begin{figure}[t!]
	\centering
	\subfloat[\scriptsize Malign image sample at 40x magnification.\label{fig:M_40-sample1}]{%
		\includegraphics[width=0.2\linewidth]{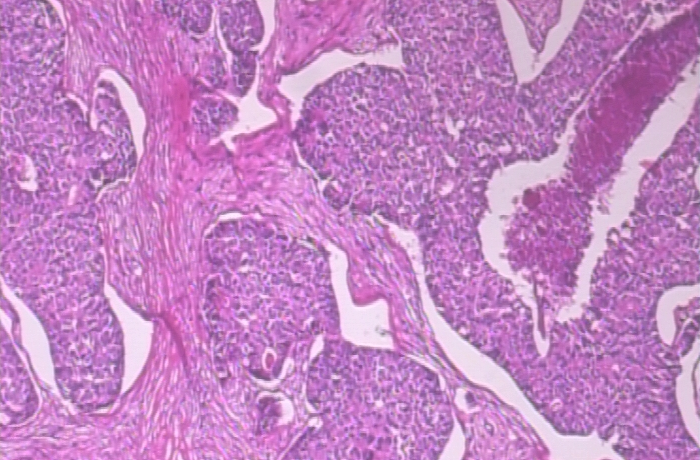}}
	\hfill
	\subfloat[\scriptsize Malign image sample at 100x magnification.\label{fig:M_100-sample2}]{%
		\includegraphics[width=0.2\linewidth]{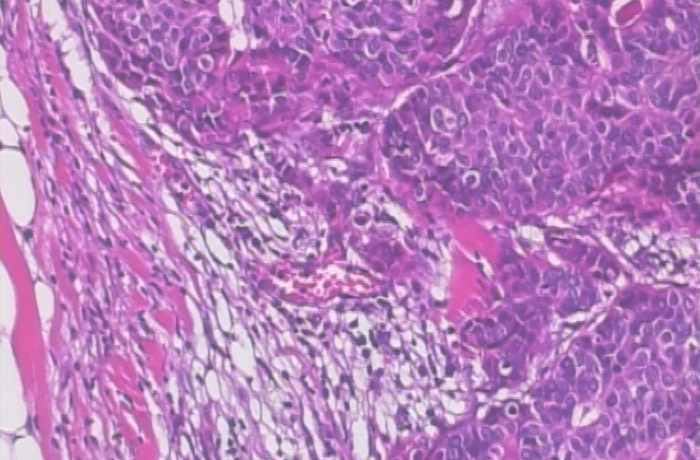}}
	\hfill
	\subfloat[\scriptsize Malign image sample at 200x magnification.\label{fig:M_200-sample3}]{%
		\includegraphics[width=0.2\linewidth]{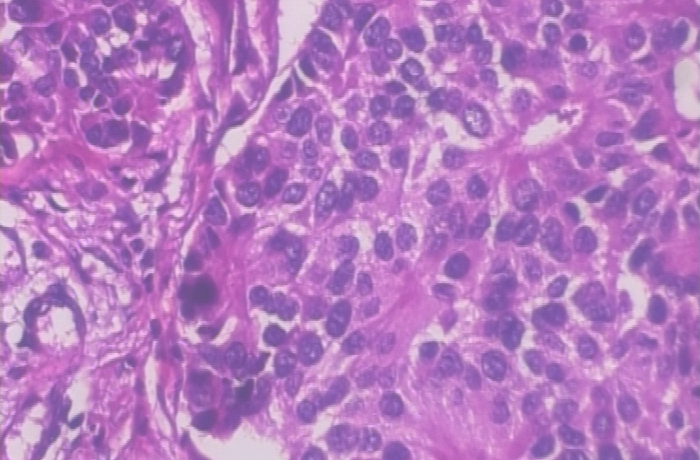}}
    \hfill
	\subfloat[\scriptsize Malign image sample at 400x magnification.\label{fig:M_400-sample2}]{%
		\includegraphics[width=0.2\linewidth]{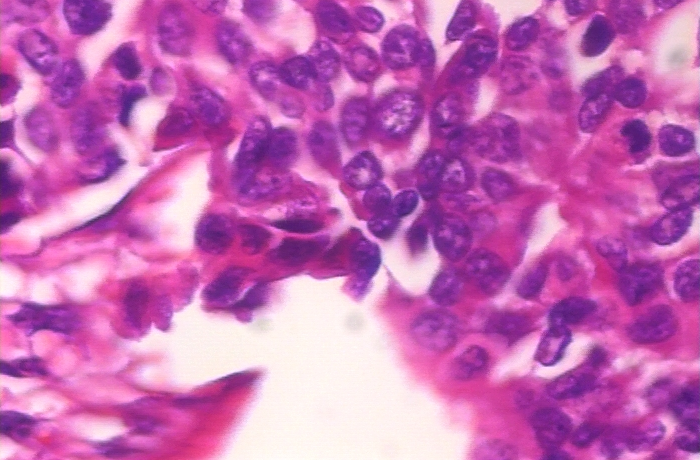}}

	\caption{\footnotesize Representative malign tissue image samples of the BreakHis dataset at different magnifications (40x, 100x, 200x, and 400x) for visual comparison.}
	\label{fig:b-image-samples}
\end{figure}

\section{Related Works}

Over the years, a significant amount of research has been conducted on breast cancer classification using both traditional machine learning and modern deep learning techniques. Among traditional machine learning methods, Support Vector Machines (SVMs) have gained popularity because of their ability to work with high-dimensional datasets. These models are paired with feature extraction methods such as Principal Component Analysis (PCA) and wavelet transforms to improve model performance by reducing the dimensionality and capturing important features \cite{YuTraditional}. Similarly, ensemble methods such as Random Forests have demonstrated robustness and high accuracy in various breast cancer diagnosis tasks by leveraging multiple decision trees \cite{MINNOOR2023429}.

In recent years, there has been a shift in focus towards deep learning approaches, which have demonstrated remarkable performance, particularly in the field of medical image analysis. Convolutional Neural Networks (CNNs) have been extensively used for classifying breast cancer in histopathological images and mammograms due to their capability to automatically learn hierarchical features from raw image data \cite{shenMammography, Srinidhi_2021}. \cite{guo2020deep, ALAHMADI2024102651} used deep learning architectures for the identification and classification of ovarian cancer subtypes.  Additionally, advanced deep learning architectures, such as ResNet, Inception, and VGG, have achieved significant success, resulting in improved accuracy in classification tasks and reduced training times for new datasets through transfer learning \cite{SANTOSBUSTOS2022101214}. These models are especially beneficial when working with limited medical imaging datasets, as they enable knowledge transfer from large-scale datasets, such as ImageNet, to the domain of medical diagnosis. Furthermore, state-of-the-art techniques involving topological data analysis (TDA) are also widely used in classification tasks ~\cite{ahmed2025topo, ahmed2023tofi, ahmed2023topological, ahmed2023topo, yadav2023histopathological, ahmed2025topological}. 

Vision-based models have gained popularity for classifying breast and ovarian cancers due to the increasing complexity of medical images, particularly in histopathology. As a result, researchers have started exploring transformer-based architectures for medical image analysis \cite{TransformerChris, dosovitskiy2020image, XuVision}. Among these, the Vision Transformer (ViT) has emerged as a powerful alternative to traditional CNNs \cite{pasupuletiViTs}. This is mainly because it can effectively represent long-distance relationships and global contextual information using self-attention mechanisms \cite{henry2022visiontransformersmedicalimaging}.
CNNs struggle to capture the global structure present in histopathological images because they rely on local receptive fields. In contrast, ViTs divide an image into a sequence of fixed-size patches and process them like word tokens in natural language processing. This allows ViTs to attend to features throughout the image, making them particularly useful for analyzing complex tissue morphology in cancer images. More applications of transfer learning and Vision Transformers in medical image analysis are explored in the following studies:~\cite{ahmed2025repvit, ahmed2025hog, ahmed2025ocuvit, ahmed2025robust, ahmed2025histovit, ahmed2025transfer}.

The paper by \cite{vitLaid} applied three versions of vision transforms for breast cancer classification at different magnification levels. One served as the baseline ViT, while the other two were modified versions. Another study employed a hybrid model called TokenMixer to analyze the BreakHis dataset \cite{LaidTokenMixer}.


\section{Method}

The following section details the methodological approach used for preprocessing image data, defining the dataset, utilizing a pre-trained Vision Transformer (ViT) model, and training it for multi-class and binary image classification. The process is systematically outlined as follows.  

In the preprocessing stage, each image is formatted as 3D tensor $\mathbf{I}_{\text{raw}}$ of dimensions $(H, W, C)$, where $H$ and $W$ denote the height and width of the image, and $C = 3$ represents the RGB color channels. The pixel values of the raw image are normalized to a range of $[0, 1]$ using the formula:  
\[
\mathbf{I}_{\text{norm}} = \frac{\mathbf{I}_{\text{raw}}}{255}.
\]  
The normalized image tensor is then permuted to match PyTorch's expected input format $(C, H, W)$ as follows:  
\[
\mathbf{I}_{\text{norm}} \rightarrow \mathbf{I}_{\text{permute}}(C, H, W).
\]  
To enable efficient training, multiple images are stacked into batches of size $B$, represented as:  
\[
\mathbf{X} = \{\mathbf{I}_1, \mathbf{I}_2, \ldots, \mathbf{I}_B\}, \quad \mathbf{y} = [y_1, y_2, \ldots, y_B],
\]  
where $\mathbf{y}$ contains the corresponding labels.  

A custom PyTorch dataset class is defined to manage images and their labels. For the $i$-th data point, the class provides access to the image $\mathbf{I}_i$ and its corresponding label $y_i$, where $y_i \in \{0, 1, 2, 3, 4\}$ for multi-class and $y_i \in \{0, 1\}$ for binary class. A train-test split is applied to the dataset to enable model training and performance assessment 
, and a \texttt{DataLoader} is used to create batches.  

Fine-tuning is performed on a pre-trained Vision Transformer (ViT) to adapt both binary and multi-class classification objectives. Each input image $\mathbf{I}(C, H, W)$ is divided into $N$ patches of size $(P \times P)$, which are flattened and projected into a latent space using learnable weights $\mathbf{W}$ and biases $\mathbf{b}$:  
\[
\mathbf{E}_i = \mathbf{W} \cdot \text{Flatten}(\text{Patch}_i) + \mathbf{b}, \quad i = 1, \ldots, N.
\]  
In order to preserve spatial relationships, positional encodings $\mathbf{p}_i$ are incorporated into the patch embeddings, resulting in: 
\[
\mathbf{z}^0 = [\mathbf{x}_{\text{cls}}, \mathbf{E}_1 + \mathbf{p}_1, \ldots, \mathbf{E}_N + \mathbf{p}_N],
\]  
where $\mathbf{x}_{\text{cls}}$ is a special classification token. The embeddings are processed through $L$ Transformer layers, each comprising multi-head self-attention mechanisms followed by feed-forward neural networks. Self-attention computes the output representation using the following formulation:  
\[
\text{Attention}(\mathbf{Q}, \mathbf{K}, \mathbf{V}) = \text{Softmax}\left(\frac{\mathbf{QK}^\top}{\sqrt{d_k}}\right) \mathbf{V},
\]  
where $\mathbf{Q}$, $\mathbf{K}$, and $\mathbf{V}$ are query, key, and value matrices, and $d_k$ is the dimensionality of the keys. Stability is ensured by incorporating residual connections and layer normalization, with the output updated as:  
\[
\mathbf{z}^{\ell+1} = \text{LayerNorm}(\mathbf{z}^\ell + \text{FFN}(\mathbf{z}^\ell)).
\]  
Finally, the classification token $\mathbf{z}^L_{\text{cls}}$ is processed by a fully connected (dense) layer to produce class probability distributions using:  
\[
\hat{\mathbf{y}} = \text{Softmax}(\mathbf{W}_{\text{cls}} \mathbf{z}^L_{\text{cls}} + \mathbf{b}_{\text{cls}}).
\]  

The training process optimizes the model parameters through the minimization of the cross-entropy loss function, defined as:  
\[
\mathcal{L} = -\frac{1}{B} \sum_{i=1}^B \sum_{c=1}^C y_{i,c} \log(\hat{y}_{i,c}),
\]  
where $C$ is the number of classes. Gradients of the loss with respect to model parameters $\theta$ are computed using backpropagation, and the model parameters are updated using the Adam optimizer according to the following procedure:  
\[
\theta_{t+1} = \theta_t - \eta \cdot \nabla_\theta \mathcal{L},
\]  
where $\eta$ is the learning rate.  

For evaluation, the model generates predictions $\hat{\mathbf{y}}$ by applying the softmax function to logits $\mathbf{z}$ and determines the predicted class as:  
\[
\hat{y}_i = \arg\max_c \hat{y}_{i,c}.
\]  
The model's accuracy is calculated as:  
\[
\text{Accuracy} = \frac{\text{Number of Correct Predictions}}{\text{Total Number of Predictions}} \times 100,
\]  
and the cross-entropy loss is reevaluated on the test set to quantify the error.  

The entire process is implemented using PyTorch, fine-tuning the Vision Transformer on a GPU when available. Training is performed for 50 epochs with a batch size of 32, and performance metrics, including loss and accuracy, are logged during both training and testing phases. The flowchart of the pre-trained ViT model is depicted in Figure~\ref{fig:flowchart}. Additionally, the detailed algorithm for our model OcuViT is provided in Appendix~\ref{alg:vit_classification}.

\begin{figure*}[t!]
    \centering
     \includegraphics[width=\linewidth]{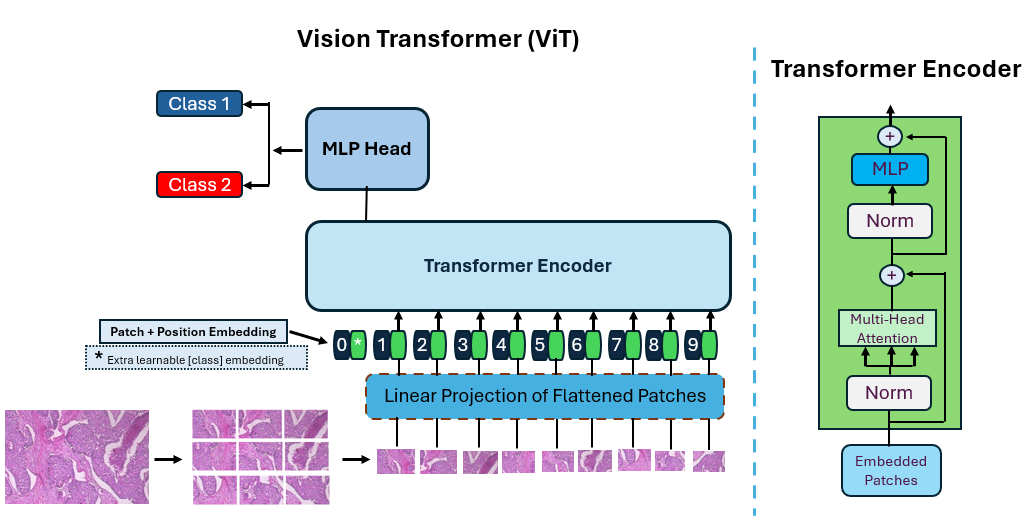}
     \caption{\footnotesize {Flowchart of ViT pretrained model:} The model design is inspired by \cite{dosovitskiy2020image}. We begin by dividing the image into fixed-size patches, linearly embedding each patch, and incorporating positional embeddings. The resulting sequence of vectors is then processed through a standard Transformer encoder. For classification, we follow the conventional method of appending an additional learnable "classification token" to the sequence.}
     \label{fig:flowchart}
 \end{figure*}

\section{Experiment}

\subsection{Datasets}
The first dataset that we used to evaluate our model is the Breast Cancer Histopathological Image Classification dataset, also known as BreakHis \cite{spanhol2016dataset}, which is one of the widely used datasets for breast cancer classification. It comprises a total of $7,909$  microscopic images of breast tumor tissue samples collected from $82$ patients at the PD laboratory. The dataset is divided into two primary categories: benign and malignant tumors. Specifically, there are $2,480$ images labeled as benign and $5,429$ labeled as malignant.
 The images were captured at four different magnification levels—40x, 100x, 200x, and 400x, and are stored in RGB format with a resolution of $700\times460$ pixels. The varying magnification levels provide diverse and detailed insights into the cellular structures. In addition to the broad classifications of benign and malignant tumors, the dataset is further classified into eight subtypes: adenosis, fibroadenoma, phyllodes tumor, tubular adenoma (benign), ductal carcinoma, lobular carcinoma, mucinous carcinoma, and papillary carcinoma (malignant). Furthermore, as the dataset comprises data from different patients, it also exhibits inter-patient variability, including differences in tissue morphology, staining techniques, and imaging conditions. These factors make BreakHis a challenging yet realistic dataset for evaluating model performance. 
 For our analysis, we used an $85:15$ split of the dataset for training and testing.

One of the other datasets we used for our analysis is the UBC-OCEAN dataset, as referenced in the previous work \cite{Fatema_2024}.
The UBC-OCEAN dataset was developed by the University of British Columbia in collaboration with Vancouver General Hospital. It is a curated collection of hematoxylin and eosin (H\&E)-stained histopathology whole-slide images (WSIs) of ovarian tissue samples.  The dataset includes five different subtypes of ovarian cancer: serous, endometrioid, clear cell, and mucinous carcinomas, as well as normal ovarian tissue samples. This makes the dataset suitable for both binary classification, distinguishing between cancerous and non-cancerous tissues, and multi-class classification across different subtypes. This dataset is organized into predefined training and testing subsets containing $31203$ training images and $3,082$ test images. We have used an $85:15$ training and test split for our analysis.

\subsection{Experimental Setup}
\noindent {\textbf No Data Augmentation:} In constrast to traditional CNNs and deep learning methods that typically rely on extensive data augmentation to address small, imbalanced datasets, our model, employs pre-trained backbones and eliminates the need for augmentation. This approach improves computational efficiency and provides robustness against minor alterations and noise in the images.
\noindent {\textbf Runtime Platform:} 
The experiments were performed on the UTA high-performance computing cluster (UTA HPC) using Intel$\circledR$ Xeon$\circledR$ E5-$2699$ processors with CentOS Linux $7.6$ as an operating environment with more than 3.5 terabytes of RAM. \\
Training was carried out for $50$ epochs with a batch size of $32$, using early stopping with a patience value of $10$. Our code is available at \footnote{\url{https://github.com/FaisalAhmed77/OcuViT}}.

\section{Results}
The results presented in this paper highlight the effectiveness of our Vision Transformer-based approach, which achieves state-of-the-art performance in the classification of breast and ovarian cancer datasets.
 
In Table \ref{tab:table1}, we present results for multiclass classification of the ovarian dataset.  We compare the accuracy of our model with the state-of-the-art model TopOC-CNN introduced by \cite{Fatema_2024}. There are various versions of the TopOC-CNN model, depending on the type of CNN architecture used and varying dimensions of the topological feature vector. Our proposed method performs well across all three performance metrics: balanced accuracy, accuracy, and area under curve (AUC).

In Table \ref{tab:table2}, we present results of different methodologies applied to the BreakHis dataset for binary classification. The assessment of these methods is based on three metrics: accuracy (Acc.), sensitivity (Sens.), and specificity (Spec.). Our proposed model performs well with the highest accuracy and sensitivity, while also achieving second-best specificity, just behind the established ViT baseline model. This comprehensive evaluation demonstrates the superiority of our approach over existing CNNs and hybrid models in the field of cancer identification.

Lastly, in Table \ref{tab:table3}, we present a comprehensive overview of our model's performance across both datasets. We illustrate our model performance using key metrics including accuracy, precision, recall, and AUC.
 
\begin{table}[h]

\caption{Accuracy results of our model compared to the TopOC-CNN model \cite{Fatema_2024} with different CNN architectures and varying dimensions of the topological feature vector (0, 64, 128, 256) on the multiclass classification of the UBC OCEAN dataset.}
\begin{tabular}{@{}llll@{}}
\toprule
\textbf{Method}  & \textbf{B.Acc} & \textbf{Acc.} & \textbf{AUC} \\
\midrule
DenseNet 121 (Vanilla-CNN)  & 65.10 & 69.08 & 89.42 \\
DenseNet 121 (CNN+64Top ) &64.62 & 68.59 & 89.37 \\
DenseNet 121 (CNN+128Top )  & \underline{67.15} & 69.34 & \underline{92.08}  \\
DenseNet 121 (CNN+256Top)  & 66.36 & \underline{69.99} & 89.62\\
EfficientNetB0 (Vanilla-CNN)  & 56.08 & 63.82 & 87.04  \\
EfficientNetB0 (CNN+64Top ) & 59.85 & 64.34 & 86.71  \\
EfficientNetB0 (CNN+128Top)  & 59.96 & 64.05 & 86.48  \\
EfficientNetB0 (CNN+256Top) & 62.67 & 67.98 & 89.14 \\
VGG16 (Vanilla-CNN) & 56.45 & 61.55 & 84.02 \\
VGG16 (CNN+64Top ) & 56.50 & 61.26 & 83.53   \\
VGG16 (CNN+128Top) & 55.00 & 62.20 & 84.33   \\
VGG16 (CNN+256Top)  & 57.63 & 63.24 & 84.24 \\
\midrule
\textbf{Proposed Model} & \textbf{88.87} & \textbf{89.64} & \textbf{98.84} \\
\bottomrule
\end{tabular}
\footnotetext[1]{Bold values indicate the highest performance for a given metric.}
\footnotetext[2]{Underlined values represent the second-best performance.}
\label{tab:table1}
\end{table} 

\begin{sidewaystable}[!htbp]
\centering
\caption{Accuracy results of our model compared to other deep learning models for binary diagnosis on the BreakHis dataset.}
\begin{tabular}{l ccc ccc ccc ccc }
\toprule
\textbf{Method} & \multicolumn{3}{c}{\textbf{40x}} & \multicolumn{3}{c}{\textbf{100x}} & \multicolumn{3}{c}{\textbf{200x}} & \multicolumn{3}{c}{\textbf{400x}} \\
\cmidrule(lr){2-4} \cmidrule(lr){5-7} \cmidrule(lr){8-10} \cmidrule(lr){11-13}
 & \textbf{Acc.} & \textbf{Sens.} & \textbf{Spec.}
 & \textbf{Acc.} & \textbf{Sens.} & \textbf{Spec.}
 & \textbf{Acc.} & \textbf{Sens.} & \textbf{Spec.}
 & \textbf{Acc.} & \textbf{Sens.} & \textbf{Spec.} \\ 
\midrule
AlexNet \cite{Alexnet, WANGBKCaps}       & 81.52 & 75.64 & 87.40 & 81.28 & 78.16 & 84.40 & 83.54 & 79.16 & 87.91 & 81.10 & 76.28 & 85.90 \\
BkNet \cite{WANGBKCaps}         & 85.61 & 84.42 & 86.80 & 86.23 & 87.19 & 85.26 & 85.37 & 80.01 & 90.74 & 84.43 & 80.00 & 88.87 \\
CapsNet \cite{sabour2017dynamic, WANGBKCaps}     & 86.95 & 86.29 & 87.61 & 89.13 & 88.30 & 89.96 & 88.75 & 86.22 & 91.28 & 88.04 & 87.56 & 88.51 \\
E-CapsNet \cite{WANGBKCaps}         & 91.67 & 90.09 & 93.25 & 93.87 & 94.36 & 93.39 & 93.34 & 93.64 & 93.04 & 92.85 & 92.69 & 93.01 \\
FE-BkCapsNet \cite{WANGBKCaps}     & 92.71 & 92.15 & 93.27 & 94.52 & 95.16 & 93.87 & 94.03 & 94.31 & 93.75 & 93.54 & 94.06 & 93.03 \\
TopOC-1 \cite{Fatema_2024}      & 90.82 & \underline{97.76} & 76.77 & 91.68 & \underline{96.00} & 82.50 & 91.05 & 96.53 & 80.00 & 88.64 & \underline{96.02} & 75.25 \\
TopOC-CNN \cite{Fatema_2024}    & \underline{94.82} & 94.06 & 92.02 & 92.48 & 91.03 & 88.08 & 93.71 & 92.79 & 90.37 & 93.22 & 92.17 & 89.20 \\
ViT \cite{dosovitskiy2020image, LaidTokenMixer}  & 93.32 & 93.45 & \underline{93.45} & \underline{95.10} & 95.03 & \underline{95.03} & 96.53 & 95.28 & 95.28 & \underline{95.38} & 94.66 & \underline{94.66} \\
TokenMixer \cite{LaidTokenMixer}   & 92.75 & 92.57 & 92.57 & 94.64 & 93.31 & 93.31 & \underline{97.02} & \underline{96.74} & \underline{96.74} & 95.11 & 94.88 & \textbf{94.88} \\\midrule
\textbf{Proposed Method}    & \bf{99.33} & \bf{99.33} & \bf{99.63}  & \bf{97.44} &  \bf{97.44}& \bf{95.15} & \bf{98.68}  & \bf{98.68} & \bf{97.48} & \bf{96.34} & \bf{96.34} & 93.68  \\
\bottomrule
\end{tabular}
\footnotetext[1]{Bold values indicate the highest performance for a given metric.}
\footnotetext[2]{Underlined values represent the second-best performance.}
\label{tab:table2}
\end{sidewaystable}

\begin{table}[!htbp]
\centering
\caption{Performance metrics of our model for BreakHis and UBC OCEAN Dataset}
\begin{tabular}{l ccccc }
\toprule
\textbf{Dataset} & \textbf{Accuracy} & \textbf{Precision} & \textbf{Recall}  &\textbf{AUC}\\ 
\midrule
BreakHis-40x      & 99.33 & 99.35 & 99.33 & 99.27 \\
BreakHis-100x      & 97.44 & 97.49 & 97.44 & 96.30 \\
BreakHis-200x      & 98.68 & 98.70 &  98.68 & 98.08 \\
BreakHis-400x      & 96.34 & 96.38 & 96.34 & 95.01 \\
UBC OCEAN      & 89.64 & 90.02 & 89.64 & 98.84 \\
\bottomrule
\end{tabular}
\label{tab:table3}
\end{table}

\section{Discussion}

The results from both the UBC OCEAN and BreakHis datasets demonstrate that the proposed model provides a significant advancement over existing CNN, capsule-based, and transformer-based architectures.  

On the UBC OCEAN dataset, our model achieved a balanced accuracy of \textbf{88.87\%}, accuracy of \textbf{89.64\%}, and an AUC of \textbf{98.84}. In comparison, the best-performing TopOC-CNN with DenseNet121 backbone achieved a maximum balanced accuracy of only $67.15\%$, accuracy of $69.99\%$, and AUC of $92.08$. This represents an improvement of more than $20\%$ in balanced accuracy and accuracy, and nearly $7\%$ in AUC. These results indicate that our approach is more effective at handling the multiclass histopathology classification problem, particularly in terms of balanced accuracy, which highlights the robustness of the model in addressing class imbalance issues that often arise in medical image datasets.  

On the BreakHis dataset, our method also demonstrated superior performance across all magnification levels. At 40x magnification, the proposed model reached an accuracy of \textbf{99.33\%}, sensitivity of \textbf{99.33\%}, and specificity of \textbf{99.63\%}, surpassing the previous best performance by more than $4\%$ in accuracy. At 100x, our model maintained a strong accuracy of \textbf{97.44\%}, with sensitivity of \textbf{97.44\%} and specificity of \textbf{95.15\%}, again outperforming models such as ViT and TokenMixer, which achieved accuracies around $95\%$. At 200x magnification, the accuracy was \textbf{98.68\%} with sensitivity of \textbf{98.68\%} and specificity of \textbf{97.48\%}, showing a clear advantage over transformer-based methods that reported accuracies close to $97\%$. Even at the highest 400x magnification, where performance often degrades, our model achieved an accuracy of \textbf{96.34\%}, outperforming TopOC-CNN and matching or exceeding state-of-the-art transformer models.  

Taken together, these findings demonstrate that the proposed method not only delivers the highest performance in terms of accuracy but also maintains excellent sensitivity and specificity, which are crucial in clinical decision-making. High sensitivity ensures that malignant cases are correctly identified, while high specificity reduces false positives, both of which are vital for reducing diagnostic errors. The consistent improvements across datasets and magnifications further highlight the generalization ability of our model.  

Overall, the results suggest that our approach effectively integrates both discriminative feature extraction and robust decision-making, providing a reliable framework for computer-aided histopathology diagnosis. This positions the proposed method as a strong candidate for real-world applications, where accuracy, sensitivity, and specificity are equally critical.





\section{Conclusion}
In this paper, we introduce a novel approach that uses Vision Transformer (ViT) for breast and ovarian cancer classification using histopathological whole-slide images. We use transfer learning with a pre-trained ViT model to improve our classification results. Our approach has demonstrated state-of-the-art classification performance across two well-known benchmark datasets: BreaKHis, for binary classification, and the UBC OCEAN dataset for multiclass classification. 
Our proposed model surpasses the performance of current CNN and Transformer-based methods in binary classification. For multi-class classification, it outperforms the existing state-of-the-art method known as TopOC-CNN. We evaluated our models using various metrics like balanced accuracy, accuracy, AUC, sensitivity, and specificity. Specifically for the BreakHis dataset, our model achieved an impressive accuracy of $99.33\%$. For the UBC OCEAN dataset, we achieved a balanced accuracy of $88.87\%$ and an accuracy of $89.64\%$, surpassing the state-of-the-art method TopOC-CNN with different backbones and varying final output layer settings. These results highlight the need to focus on Vision Transformer-based architectures, which are capable of capturing complex long-range dependencies and thereby improving diagnostic precision in cancer detection. 











\newpage
\begin{appendices}

\section{Algorithm of the Proposed Model}

\begin{algorithm}
\SetAlgoNlRelativeSize{0} 
\DontPrintSemicolon 
\caption{Image Classification Using Proposed Model}
\label{alg:vit_classification}

\KwIn{Image Dataset $I_i$ with corresponding label $y_i$, $\mathcal{D} = \{(\mathbf{I}_i, y_i)\}_{i=1}^{N}$, pre-trained ViT, epochs $E$, and number of classes $C$}
\KwOut{Trained proposed model and classification metrics}
\textbf{Preprocessing:} \\
\For {$i \gets 1$ \KwTo $ N$ }{
$\mathbf{I}_i \gets \mathbf{I}_i / 255$ \hfill $ \vartriangleright$ Normalize images\\
$\mathbf{I}_i \rightarrow \mathbf{I}_i (C, H, W)$ \hfill $\vartriangleright$ Permute image dimensions\\
$\mathcal{D}=\mathcal{D}_{train}\bigcup \mathcal{D}_{test}$ \hfill $\vartriangleright$ Split dataset into training and testing sets
}

\textbf{Model Initialization:} \\
Load pre-trained ViT model\;
Modify the classification head for $C$ classes\;

\For{$\text{epoch} \gets 1$ \KwTo $E$}{
    \textit{model=model.train()} \hfill $\vartriangleright$ Set model to train mode\\
    \ForEach{\text{batch } $(\mathbf{X}, \mathbf{y})$ in $\mathcal{D}_{train}$}{
        $\hat{\mathbf{y}} \gets \text{model}(\mathbf{X})$ \hfill $\vartriangleright$ Forward pass\\
        $\mathcal{L} = \text{CrossEntropyLoss}(\hat{\mathbf{y}}, \mathbf{y})$ \hfill $\vartriangleright$ Compute loss \\
        Backpropagate loss and update parameters\;
    }
    Compute training loss and matrices\;

    \textbf{Testing Phase:} \\
    \textit{model=model} \hfill $\vartriangleright$ Set model to evaluation mode\\
    \ForEach{\text{batch } $(\mathbf{X}, \mathbf{y})$ in $\mathcal{D}_{test}$}{
        $\hat{\mathbf{y}} \gets \text{model}(\mathbf{X})$ \hfill $\vartriangleright$ Forward pass \\
        Compute loss and matrices\;
    }
    Log test loss and matrices\;
}

\Return{Trained model and performance metrics}
\end{algorithm}

\end{appendices}

\clearpage

\bibliographystyle{elsarticle-num-names}

\bibliography{refs}

\end{document}